# Y358 against Y123 structural phase in a Y-based superconductor

S Gholipour[1], V Daadmehr[1], A T Rezakhani[2], H Khosroabadi[2], F Shahbaz Tehrani[1] and R Hosseini Akbarnejad[1]

1. Magnet and Superconducting Research Lab, Department of Physics, Alzahra University, Tehran 19938, Iran

2. Department of Physics, Sharif University, Tehran, Iran

Email: daadmehr@alzahra.ac.ir

**Abstract**

We report an experiment in which two distinct superconducting phases $YBa_2Cu_3O_7$ (Y123) and $Y_3Ba_5Cu_8O_{18}$ (Y358) coexisted. This enabled us to characterize the recently introduced Y358 phase in contrast to the Y123 phase, thus to resolve some discrepancies reported in associated properties of Y358. Specifically, our experiment indicates the transition temperature $T_C^{mid}$=105K and 94K for Y358 and Y123, respectively, and that Y358 has three $CuO_2$ planes and two CuO chains, with Pmm2 symmetry and lattice parameter (a,b,c)=(3.845,3.894,31.093)Å, in agreement with density functional theory predictions for this specific structure.

## 1. Introduction

Since the discovery of high-$T_C$ superconductors (HTSCs) [1], increasing critical temperature and current of superconductors has been the aim of intensive research. As a result, it has been shown that YBCO compounds exhibit a high critical temperature [2]. This has spurred a lot of interest in such compounds for their applications in various areas [3].

In the YBCO family, the structural differences, the number of $CuO_2$ planes and CuO chains lead to various physical properties and critical temperatures. For example, Y123 compounds have two $CuO_2$ planes and one CuO chain, exhibiting a $T_C$ in the 92-94K range; whereas, Y124 compounds have one $CuO_2$ plane and one double CuO chain, featuring a $T_C \approx$ 80K [4].

Recently, a new member of the YBCO family, Y358, has been found to feature a $T_C \geq 100K$ [5]. Y358 has five $CuO_2$ planes and three CuO chains, grouped in two sets: three $CuO_2$ planes separated by BaO layers from the other two $CuO_2$ planes. In this sense, it is also different from similar cuprates.

The over 100K critical temperature of Y358 compounds has stimulated more detailed investigations on their structure. For example, density functional theory calculations (based on the structure shown in Figure 1) have also verified the suggested structure of Y358 compounds [6]. More specifically, these calculations showed that the hole concentration in the chains of Y123, Y247, Y124, and Y358 plays an important role in why Y358 and Y247 feature higher $T_C$ than the rest: the improvement in $T_C$ has been attributed to lower number of holes in the CuO chains of Y358 and Y247 relative to Y123 and Y124.

In a separate experiment, however, another group reported synthesis of a Y358 compound with $T_C^{on} \approx 94K$, with lattice parameters almost identical to those of Y123 [7]. This result is in conflict with that of Refs. [5, 6]. Since then, two more groups have attempted to characterize Y358 [8, 9, 10]; one obtaining lattice parameters similar to Y123 [7, 8, 9], while the other group has reported lattice parameters similar to those of Refs. [5, 6], but with a different symmetry group [10].

In this work, we aim to resolve the above discrepancies regarding Y358. We synthesized a sample which has both Y123 and Y358 structural phases. The existence of both Y123 and Y358 structures in our sample has been verified by the X-ray diffraction supplemented with the MAUD software calculations. This co-existence of the phases has enabled us to differentiate them clearly by characterizing them with associated lattice parameters and symmetry groups. As a result, we explicitly verify the conclusion of Ref. [5] regarding Y358 compounds.

**2. Experiment**

The sol-gel technique is one of the best procedures for fabricating HTSCs due to higher homogeneity and purity of raw materials as well as lower temperature of preparation. As a result, we synthesized samples of Y358 compound by the sol-gel technique. Appropriate stoichiometric compositions of Y:Ba:Cu=3:5:8 of $Cu(NO_3)_2 \cdot 3H_2O$ (99.5%), $Ba(NO_3)_2$ (99%), $Y(NO_3)_3 \cdot 5H_2O$ (99%) powders were weighed at $10^{-5}$gr precision (AND®), solved in deionized water. We added these solutions to 0.5M solution of citric acid ($C_6H_7O_8$) with 1:1 mol ratio for nitrates:citric acid. The pH of the final solution was adjusted to 6.8 by ethylendiamine [11]. This aqueous solution was allowed to evaporate on a stirrer hot plate maintaining the solution temperature at 83±1°C. The prepared gel was heated at 520°C for 2h in air atmosphere and cooled to room temperature (≈25°C) to eliminate O-H groups. Next, the resultant powder was calcined in air at 790°C for 12h, and cooled to 300°C in 8h. Calcination was repeated twice in order to remove the remaining nitrates from the specimen. The powder was pressed into pellets with 13mm diameter under 50Kg/cm² pressure. To prepare a special sample in which both of Y123 and Y358 phases coexist, we fabricated five samples with different sintering diagrams. In our experiment, only two of these samples exhibited the desired phase coexistence. We choose, from these two samples, the one for which resistivity-temperature diagram showed a better distinguishability of the two phases. Figure 2 depicts the details of the sintering process for the desired sample.

The sintering and oxygenating are the most important steps in the preparation of HTSCs. Thus the prepared pellets were sintered through a specific procedure (shown in Figure 2) at 860˚C for 24h in an oxygen atmosphere. The samples then were cooled to 260˚C in oxygen flow for 24h. We characterized crystal features such as the crystal morphology, lattice parameters, Miller indices, crystallite size, and dominant and/or undesirable phases of our samples (the two featuring both phases) by the X-ray diffraction pattern (XRD) [Philips® PW1800 with Cu $\kappa_\alpha$ radiation with λ=1.542A˚ (2θ range of 4-90˚)]. The mean grain size of the samples was examined by Hitachi® S4160 Field Emission Scanning Electron Microscopy (FE-SEM). A four-probe technique was used for transport measurements. The size of the samples was about 9×6×1mm$^3$, and the applied DC current (Lake Shore®-120) was 10mA. The temperature was monitored by a Lutron®-TM917 (with 0.01K resolution). The electrical resistance of the sample at room temperature was less than 0.7Ω and the voltmeter accuracy was in the μV range.

Note that the following discussions are on the results of the experiment with the desired sample.

## 3. Results and discussion

X-ray patterns are used to obtain a wide range of physical properties of synthesized samples, such as crystallinity, stress, texture, and crystallite size. The main application of XRD in HTSCs is to identify underlying phases.

Having this utility, thus, we also employed the XRD pattern of our sample, refined by the Materials Analysis Using Diffraction (MAUD) software based on the Rietveld analysis, to determine associated crystal structure and phases. Figure 3 shows the measured and calculated XRD patterns, their difference, and the designated Miller indices. Based on the Rietveld analysis, we obtained that the sample comprised of 14%/86% volume fraction of the Y358 and Y123 phases, with Pmm2 and Pmmm symmetry, respectively. Note from the XRD pattern that other undesirable phases are negligible in the synthesized compound.

The obtained symmetries were in agreement with Refs. [5, 6]. In addition, from the refinement process, we obtained the lattice parameters (a,b,c)=(3.845,3.894,31.0931)Å – with the unit cell volume V=465.53Å$^3$ – for the Y358 phase. We notice that our values of 'a' and 'b' are in agreement with Refs. [5, 6, 7, 8, 9,10]; whereas 'c' agrees with Refs. [5, 6, 10], it is three times the value reported in Ref. [7,8, 9]. Similarly, for the Y123 phase we obtained (a,b,c)=(3.84,3.89,11.66)Å, with unit cell volume

$V=174.172 Å^3$. The orthorombicity of the sample's structure has been obtained 0.6 through the relation 100(b-a)/(b+a). The twin peaks at 46.68° and 47.38° indicate an orthorhombic structure for both phases.

Tables 1, 2 and 3, respectively, show the result of the refined atomic positions and occupancy in the Y358 and Y123 phases. From these tables (especially the occupancy of the atoms, mentioned in the highlighted columns), it is evident that all ions are located properly in the sites suggested in Refs. [5, 6].

The FE-SEM image (Figure 4) shows that for Y358, the grain size is ≈83nm (i.e., in nano order). However, this nano order grain size does not decrease $T_C$ significantly since, due to the sintering process, the grains are glued together and form a bulk [12]. The crystallite size has been determined by Scherrer's formula, $D_{hkl}=Kλ/βcosθ$, and obtained to be 21± 2nm, which is naturally smaller than the grain size obtained from the FE-SEM image.

We have also reconfirmed the coexistence of two phases in our sample by the electrical resistivity-temperature (ρ-T) measurement. Figure 5 shows the ρ-T diagram in the 77-190K range. Two critical temperatures can be discerned from this plot: $T_C≈94K$, associated to the Y123 phase, and $T_C≈105K$, which we associate to the remaining phase, Y358, that is a $T_C^{mid}(Y358) - T_C^{mid}(Y123)=11K$ in $T_C$s difference, which is larger than the resolution of our thermometer. The $T_C(Y358)≈105K$ is in fact in good agreement with Refs. [5, 6]. Note also that the ρ-T plot indicates an insulating character for the sample in 120-188K range, whereas in the 95-115K range a metal-like behavior, with $dρ/dT=0.032$ mΩ.cm/K, is evident.

## 4. Conclusions

The YBCO family of superconductors has been shown to feature relatively high $T_C$s. A recently introduced member of this family, Y358, has been argued to exhibit $T_C≥100K$. Characterization of this phase has been the subject of some recent investigations. However, in the recently reported experiments, there are some discrepancies in distinguishing the Y358 from the Y123 phase.

Our objective here has been to resolve such discrepancies conclusively. In doing so, we prepared samples in which both of the phases coexisted. We verified existence of both phase via X-ray diffraction experiment supplemented with standard theoretical refinements.

This analysis as well as various other experiments (such as resistivity vs. temperature) enabled us to attribute two distinct superconducting phases, associated with Y123 and Y358 phases, and obtain their corresponding properties such as lattice parameters. We observed that our results are in good agreement with one of the reported experiments, thus clearly identifying the characteristics of Y358 superconductors.

**Acknowledgement**

The authors acknowledge the Iranian Nano Technology Initiative Council and Alzahra University for partial financial support.

**References**

[1] Bednorz J G and Mueller K A 1986 Possible high $T_C$ superconductivity in the Ba-La-Cu-O system *Z. Phys. B* **64** 189-193.

[2] Wu K, Ashburn J R, Torng C J, Hor P H, Meng R L, Gao L, Huang Z J, Wang Y Q and Chu C W 1987 Effect of compaction on the superconducting transition of $YBa_2Cu_3O_{9-y}$ compound *Phy. Rev. Lett.* **58** 908.

[3] Walter H, Delamare M P, Bringmann B, Leenders A and Freyhardt H C 2000 Melt-textured YBaCuO with high trapped fields up to 1.3 T at 77 K *J. Mater. Res.* **15** 1231.

[4] Marsh P, Fleming R M, mandich M L, Desantolo A M, Kwo J, Hong M and Martinez-Miranda L J 1988 Crystal structure of the 80 K superconductor $YBa_2Cu_4O_8$ *Nature* **334** 66.

[5] Aliabadi A, Farshchi Y A and Akhavan M 2009 A new Y-based HTSC with $T_C$ above 100 *Physica C* **469** 2012-14.

[6] Tavana A and Akhavan M 2010 How $T_C$ can go above 100 K in the YBCO family *Eur Phys. J. B* **73** 79-83.

[7] Udomsamuthirun P, Kruaehong T, Nikamjon T and Ratreng S 2010 The new superconductors of YBaCuO materials *J. Supercond. Nov. Magn.* **23** 1377.

[8] U. Topal, M. Akdogan, H. Ozkan, Electrical and Structural Properties of $RE_3Ba_5Cu_8O_{18}$ (RE = Y, Sm and Nd) Superconductors, J. Supercond. Nov. Magn. DOI 10.1007/s10948-011-1176-7 (2011).

[9] U. Topal and M. Akdogan, The role of Oxygenation on Superconducting Properties of $RE_3Ba_5Cu_8O_{18}$ (RE = Y, Sm and Nd) Compounds, J. Supercond. Nov. Magn. DOI 10.1007/s10948-011-1285-3 (2011).

[10] Ayaş A O, Ekicibil A, Çetin S K, Coşkun A, Er A O, Ufuktepe Y, Fırat T and Kıymaç K 2011 The structural, superconducting and transport properties of the compounds $Y_3Ba_5Cu_8O_{18}$ and $Y_3Ba_5Ca_2Cu_8O_{18}$ *J. Supercond. Nov. Magn.*10.1007/s10948-011-1192-7.

[11] Barekat Rezai S 2007 M. Sc. thesis (in Persian), Alzahra University.

[12] Xu X L, Guo J D, Wang Y Z and Sozzi A 2002 Synthesis of nanoscale superconducting YBCO by a novel technique *Physica C* **371** 129-132.

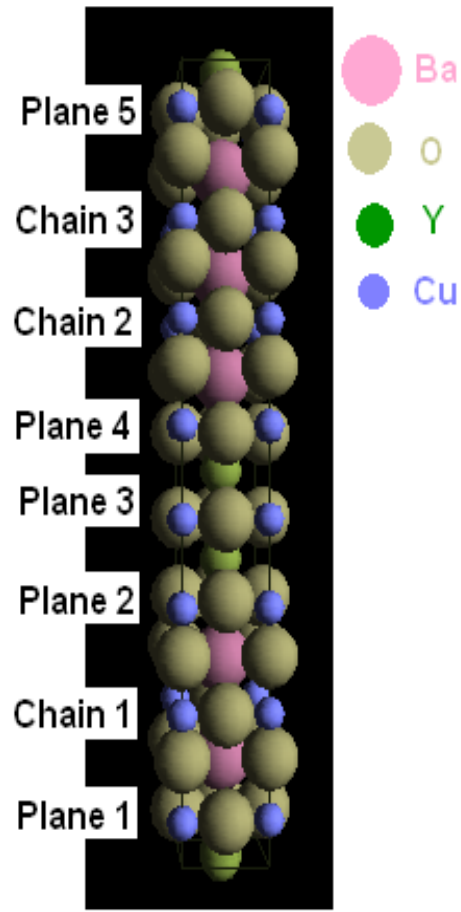

**Figure 1.** (Colour online) Crystal structure of Y358 with five CuO$_2$ planes 1 to 5 and three CuO chains 1 to 3 [5, 6].

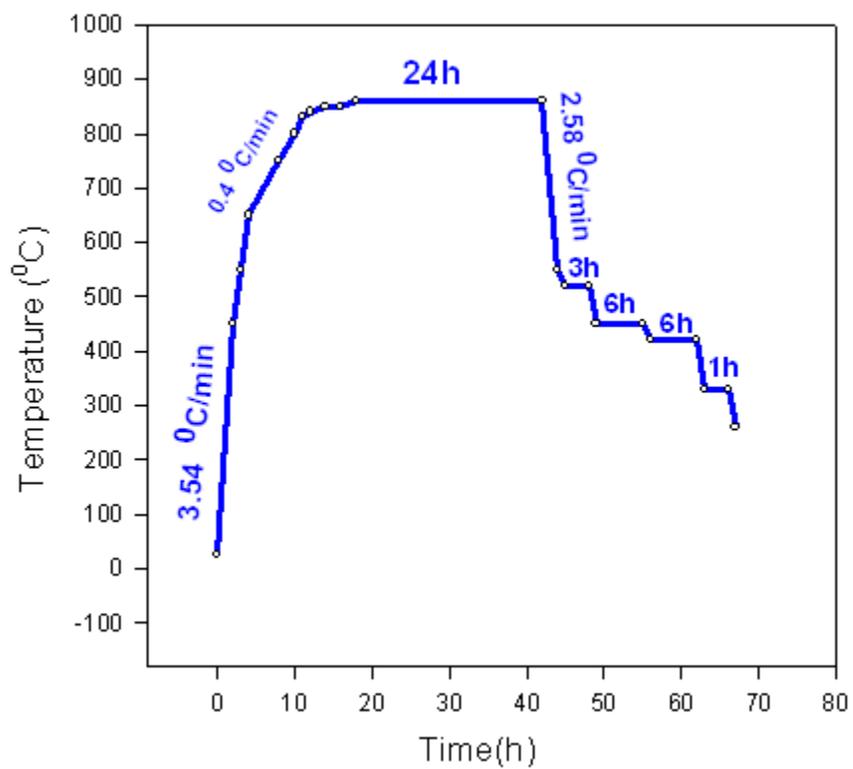

**Figure 2.** The sintering process of the $Y_3Ba_5Cu_8O_{18}$ sample.

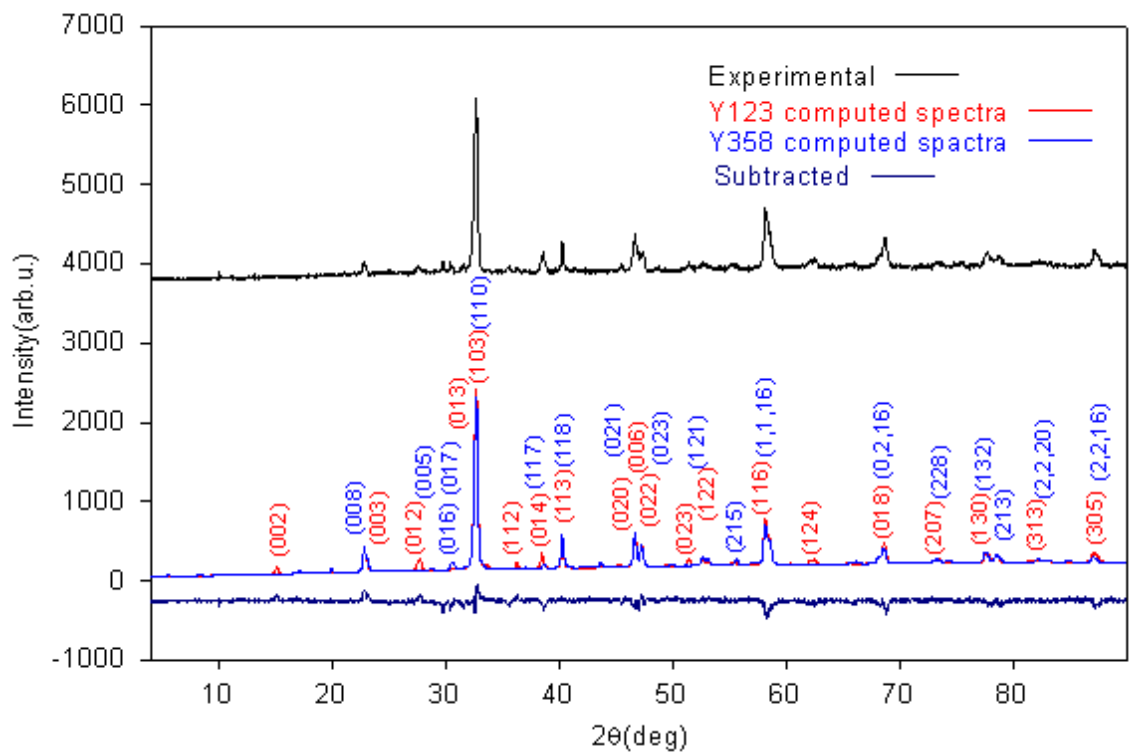

**Figure 3.** (Colour online) XRD patterns of the sample.

**Table 1.** The refinement of atomic positions and occupancy of atoms in the $Y_3Ba_5Cu_8O_{18}$ phase.

| atoms | X(A°) | Y(A°) | Z(A°) | occupancy |
|---|---|---|---|---|
| Y(1) | 0.49 | 0.49 | 0 | 0.99 |
| Y(2) | 0.49 | 0.49 | 0.37 | 1.00 |
| Y(3) | 0.5 | 0.5 | 0.48 | 1.00 |
| Y(4) | 0.49 | 0.5 | 0. 99 | 1.00 |
| Ba(1) | 0.49 | 0.49 | 0.12 | 0.99 |
| Ba(2) | 0.49 | 0.49 | 0.24 | 0.99 |
| Ba(3) | 0.5 | 0.5 | 0.61 | 0.99 |
| Ba(4) | 0.5 | 0.5 | 0.74 | 0.99 |
| Ba(5) | 0.49 | 0.49 | 0.87 | 0.99 |
| Cu(1) | 0 | 0 | 0.05 | 1.02 |
| Cu(2) | 0 | 0 | 0.18 | 1.00 |
| Cu(3) | 0 | 0 | 0.32 | 1.00 |
| Cu(4) | 0 | 0 | 0.43 | 1.00 |
| Cu(5) | 0 | 0 | 0.54 | 0.99 |
| Cu(6) | 0 | 0 | 0.67 | 0.99 |
| Cu(7) | 0 | 0 | 0.8 | 0.99 |
| Cu(8) | 0 | 0 | 0.93 | 0.97 |

**Table 2. The refinement of atomic positions and occupancy of atoms in the $Y_3Ba_5Cu_8O_{18}$ phase.**

| atoms | X(A°) | Y(A°) | Z(A°) | occupancy |
|---|---|---|---|---|
| O(1) | 0.5 | 0 | 0.04 | 0.97 |
| O(2) | 0 | 0.5 | 0.05 | 1.02 |
| O(3) | 0 | 0 | 0.12 | 0.87 |
| O(4) | 0 | 0.49 | 0.18 | 1.00 |
| O(5) | 0 | 0 | 0.25 | 0.74 |
| O(6) | 0.5 | 0 | 0.32 | 1.00 |
| O(7) | 0 | 0.5 | 0.31 | 1.00 |
| O(8) | 0.5 | 0 | 0.43 | 0.99 |
| O(9) | 0 | 0.5 | 0.43 | 1.00 |
| O(10) | 0.49 | 0 | 0.54 | 1.01 |
| O(11) | 0 | 0.49 | 0.54 | 1.00 |
| O(12) | 0 | 0 | 0.61 | 1.00 |
| O(13) | 0 | 0.49 | 0.67 | 1.00 |
| O(14) | 0 | 0 | 0.74 | 0.99 |
| O(15) | 0 | 0.53 | 0.79 | 0.99 |
| O(16) | 0 | 0.49 | 0.86 | 0.99 |
| O(17) | 0.5 | 0 | 0.93 | 0.79 |
| O(18) | 0 | 0.49 | 0.93 | 0.99 |

**Table 3.** The refinement of atomic positions and occupancy of atoms in the $YBa_2Cu_3O_7$ phase.

| atoms | X(A°) | Y(A°) | Z(A°) | occupancy |
|---|---|---|---|---|
| Y(1) | 0.49 | 0.49 | 0.49 | 0.99 |
| Ba(1) | 0.49 | 0.5 | 0.183 | 0.99 |
| Ba(2) | 0.49 | 0.5 | 0.183 | 0.99 |
| Cu(1) | 0 | 0 | 0 | 1.01 |
| Cu(2) | 0 | 0 | 0.354 | 1.00 |
| O(1) | 0 | 0 | 0.157 | 1.00 |
| O(2) | 0.46 | 0 | 0.377 | 0.98 |
| O(3) | 0 | 0.49 | 0.378 | 1.00 |
| O(4) | 0 | 0.49 | 0 | 0.99 |

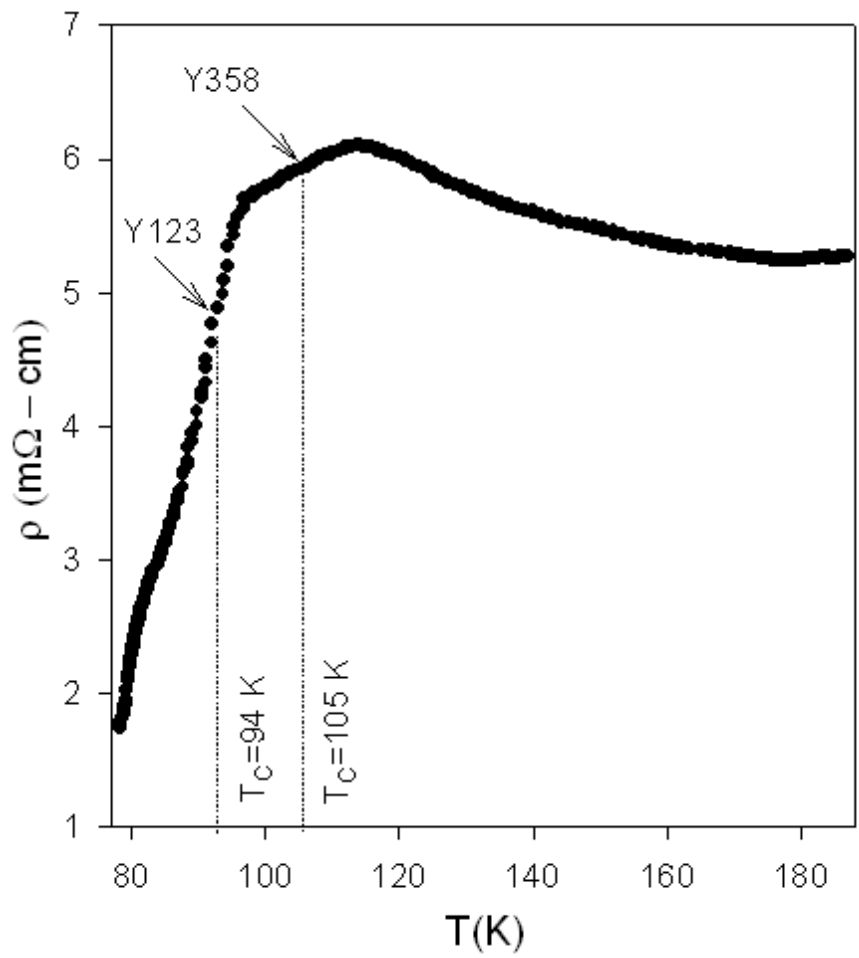

**Figure 4.** The electrical resistivity of the $Y_3Ba_5Cu_8O_{18}$ sample versus temperature.

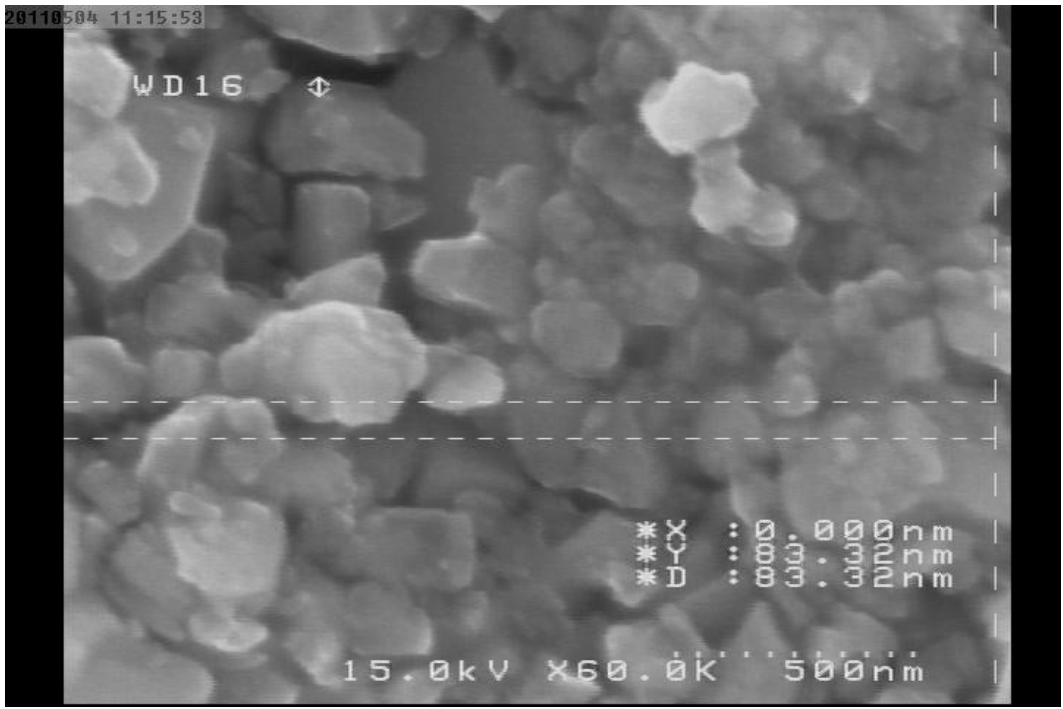

**Figure 5.** The FE-SEM image of the $Y_3Ba_5Cu_8O_{18}$ sample.